\documentclass[aps,prl,twocolumn,superscriptaddress,aps]{revtex4-1}

\usepackage{graphicx}
\usepackage{dcolumn}
\usepackage{bm}
\usepackage{amsmath}
\usepackage{bbding}
\usepackage{pifont}
\usepackage{amsfonts}
\usepackage{caption2}
\usepackage{float}
\usepackage{graphics}
\usepackage{subfigure}
\usepackage{booktabs}
\usepackage{placeins}
\usepackage{appendix}
\usepackage{flushend}
\usepackage{lineno}

\begin{document}
\modulolinenumbers[10]

\title{Predicting Streamer Discharge Front Splitting by Ionization Seed Profiling}

\author{Yujie Zhu}
\affiliation{
Department of Electrical Engineering, Tsinghua University, Beijing 100084, China
}

\author{Xuewei Zhang}
\thanks{The first two authors contributed equally}
\affiliation{
College of Engineering, Texas A$\&$M University-Kingsville, Kingsville, Texas 78363, USA
}

\author{Chijie Zhuang}
\affiliation{
Department of Electrical Engineering, Tsinghua University, Beijing 100084, China
}

\author{Rong Zeng}
\affiliation{
Department of Electrical Engineering, Tsinghua University, Beijing 100084, China
}

\author{Jinliang He}
\email{Corresponding author: hejl@tsinghua.edu.cn}
\affiliation{
Department of Electrical Engineering, Tsinghua University, Beijing 100084, China
}

\begin{abstract}
Previous studies of streamer discharge branching mechanisms have mainly been generative other than predictive. To predict or even control branching, a reliable connection between experimental conditions and streamer branching needs to be established. As an important step toward the goal, in this work, a 2D deterministic model of negative streamers in air is numerically solved with the ionization seeds assumed as the superposition of Gaussians. The ``indicative profiles approach'' developed here can consistently relate the change in a quantitative measure of geometrical irregularity of the seed profiles with specific electron densities to the emergence of front splitting of streamer discharges under various voltages, seed characteristic sizes, and preionization levels. The results of this study could inform experiments to identify and clarify streamer branching mechanisms.

\end{abstract}

\maketitle

Streamers refer to a class of electrical discharges propagating as weakly-ionized, cold plasma filaments driven by the strong electric fields at their fronts~\cite{Raizer}. Due to the unique properties of cold plasma, streamer discharges have been extensively applied to various technological fields such as decontamination and sterilization~\cite{Winands}, material processing~\cite{Oehrlein}, plasma medicine~\cite{Fridman}, aerodynamic flow control~\cite{Moreau}, and ignition and combustion~\cite{Starikovskiy}. Streamers are also the building blocks of many natural phenomena, from the inception and propagation of lightning flashes during a thunderstorm~\cite{Dubinova,Zhang}, to other less commonly known discharges in upper atmosphere like red sprites and blue jets~\cite{Shneider1,Shneider2}. In the last few years, although significant advancements have been made to elucidate physical mechanisms (e.g., ~\cite{Nijdam2010,Nijdam2016}), explore new phenomena (e.g., ~\cite{Liu2012,Heijmans2013,Nijdam2014}), and develop numerical models (e.g., ~\cite{Zhuang1,Zhuang2,Dujko2013,Luque2014,Teunissen2017}) of streamer discharges, there are still some fundamental questions of streamers that need further clarification, including, for instance, the interaction among a cluster of streamers and its effects~\cite{Luque2008,Merging2012,Milikh2016,Liu2017}. 3D simulations of the interation of two streamers have been reported in~\cite{Luque2008,Liu2017}, which discussed the two competing processes: electrostatic repulsion and tendency to merge due to nonlocal photoionization. Given the high computational costs of 3D simulations, studies have also been conducted in 2D Cartesian coordinates to derive the merging criteria for two interating streamers under various conditions~\cite{Merging2012} and show the slower propagation in the case of multiple streamers neglecting photoionization~\cite{Milikh2016}. These works assume that there is well-defined initial separation between streamers, while in reality the interacting streamers most likely emerge or branch from the same discharge front. Therefore it is necessary to illuminate the front splitting process which sets stage for the subsequent streamer interactions. This intersects with another challenge in streamer discharge physics, i.e., the mechanisms of branched pattern formation~\cite{Levko2017}.

The complex, branched patterns of electrical discharge can be generated by fractal growth models such as Dielectric Breakdown Model (DBM,~\cite{DBM}) and applied to practical engineering fields like lightning protection~\cite{Zhang1,Zhang2}. This approach, however, does not address the physical causes and processes of branching. Analyses of simplified, deterministic streamer models neglecting photoionization~\cite{MA2002,MA2005,Ebert2008,RU2013} indicate that branching is a manifest of Laplacian instability which involves a positive feedback of local curvature increase on the discharge front. It has been shown that branching follows the flattening of discharge front when the electric field maxima move away from the tip~\cite{Liu2015}. Photoionization, expected to offset or delay Laplacian instability in the deterministic branching mechanism, could be very important and sensitive in our case of the very early stage of discharge front splitting. To see this, one must solve the full model numerically. On the other hand, the numerical modeling approach has helped to identify several branching mechanisms of stochastic nature. For example, based on fluid models of streamer, it has been demonstrated that fluctuations in medium density~\cite{Papageorgiou} or charge carrier density~\cite{pre84} can result in branching. The development of particle and hybrid models (e.g., in~\cite{Li,TE2016}) enables simulations of streamer branching in which the randomness of electron collisions is modeled using Monte Carlo method and additional inhomogeneities are inherited from sampling the distributions of particle initial states. These stochastic factors generally accelerate streamer branching by creating geometrical irregularities of the discharge front which may then overcome the ``mixing'' effect of photoionization and undergo amplification via Laplacian instability. It would be beneficial to know more details on the types and features of these irregularities that cause streamer branching, which could enable the prediction of branching from known geometrical features and experimental conditions, as well as the verification of branching mechanisms in specifically-designed experiments.

Due to the strong nonlinearity and computational complexity of the problem, little work has been devoted to quantitative characterization of the geometrical irregularity and its consistent connection with streamer branching. This Letter, based on systematic simulations of the propagation, branching, and interaction of anode-directed streamers in air, demonstrates an approach that reliably relates the shape of ionization seed profiles to the emergence of front splitting. Traditionally, to simulate streamer discharge using fluid models, it is assumed that there exists an initial ionization zone (seed) where the spatial distribution of electron density is Gaussian~\cite{MA2002,probing,comparing}. In the context of lightning, hyperbolic tangent function has been used to describe seed densities~\cite{Liu2015}. However, the simple geometry of ionization seed profiles in the previous studies (usually part of an ellipsoid centered at the peak electric field point) may deviate noticeably from realistic scenarios like those shown in Fig.\ 1(a), which in this work will be approximated as the superposition of two Gaussian ionization seeds. When they are far apart, we can observe the interaction between two streamers; when they heavily overlap, we can study how irregularities in seed profile lead to streamer front splitting. A scrutiny of the latter also bears importance in providing insights into streamer discharges from complex or engineered electrode surfaces, which in turn might inform the design of experiments to identify or differentiate the aforementioned branching mechanisms.

\begin{figure}[H]
\centering
\subfigure{\includegraphics[width=250 bp]{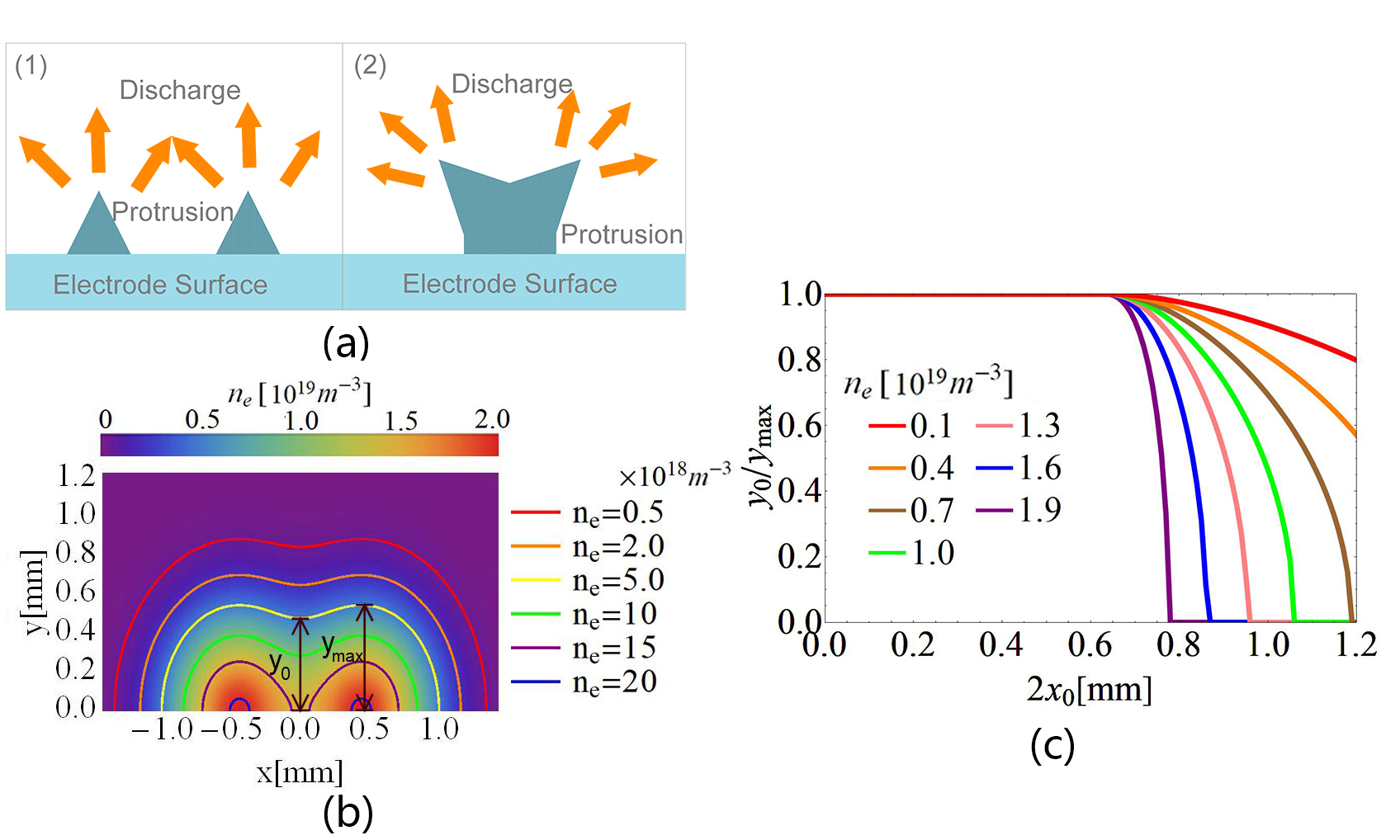}}
\caption{\footnotesize (a) Examples of rough electrode surfaces that may introduce geometrical irregularities in ionization seed profiles: (1) neighboring protrusion pairs, (2) two sharp edges of a tip; (b) The profiles (contours) of an ionization seed with the distance between two Gaussian centers $2x_{0}=0.95~$mm and various electron densities; (c) The relationship between $y_{0}/y_{max}$ and $2x_{0}$ for different electron density profiles.}
\end{figure}

Our simulation framework is as follows. In 2D Cartesian coordinates, a planar anode (applied voltage $+U$) is located at $y=10~$mm and a planar cathode (grounded) at $y=0$. The initial ionization seed on the cathode surface is set as follows: $n_{e}(x,y)|_{t=0}=n_{p}(x,y)|_{t=0}=n_{0}\exp[-(\frac{x-x_{0}}{\sigma_{x}})^2-(\frac{y}{\sigma_{y}})^2]+n_{0}\exp[-(\frac{x+x_{0}}{\sigma_{x}})^2-(\frac{y}{\sigma_{y}})^2]$, where $n_{e}$ and $n_{p}$ are the number densities of electrons and positive ions, $\pm x_{0}$ are the $x$ coordinates of the two Gaussian centers, $n_{0}=2\times 10^{19}$ $\text{m}^{-3}$ indicates the preionization level, and $\sigma_{x}$ and $\sigma_{y}$ are the characteristic sizes of each Gaussian in $x$ and $y$ directions. Fig. 1(b) plots the ionization seed profiles with $\sigma_{x}=\sigma_{y}=0.45$ mm and $2x_{0}=0.95~$mm. As the corresponding electron density increases, the profile gradually shrinks from one (approximate) semicircle to two semicircles. For any given electron density, if raising $2x_{0}$, the profile also has a similar transition. Two important quantities during this process are: the $y$ coordinate of the point with $x=0$ on the profile, denoted by $y_{0}$, and the maximum $y$ value of all points on the profile, denoted by $y_{max}$. Fig. 1(c) shows the ratio $y_{0}/y_{max}$ as functions of $2x_{0}$ for various electron densities of the profile. With $2x_{0}$ increasing from 0, $y_{0}/y_{max}$ starts decreasing from 1 at $2x_{0}=0.6\sim 0.8~$mm and eventually turns to 0 when the profile has split into two separate lobes. The lower the electron density, the slower the decrease of the ratio and the larger the critical $2x_{0}$ at which $y_{0}/y_{max}$ just reaches 0. To explore the implications of these critical values of $2x_{0}$ and the associated electron densities, this work uses COMSOL Multiphysics Plasma Module to model and simulate the streamer dynamics. The governing equations of the streamer model are based on the drift-diffusion-reaction of electrons and positive and negative ions coupled with electrostatic field~\cite{PRE}. We study streamers under standard atmospheric pressure in air (simplified as the \text{$N_{2}/O_{2}$} mixtures with the volume fraction ratio $8:2$). Following~\cite{PRE}, 7 particles (electrons $e$, $N_{2}^{+}$, $O_{2}^{+}$, $N_{4}^{+}$, $O_{4}^{+}$, $O_{2}N_{2}^{+}$, $O_{2}^{-}$) and 15 reactions are taken into account. The calculation of the photoionization term is based on the three-term exponential Helmholtz model developed in~\cite{17} and more recently used by~\cite{PIC}. In the simulations, the size of the problem domain is 1.0$\times$1.0 cm$^{2}$, the finest mesh size (on and near the symmetric axis $x=0$) is set as $0.2~$\text{$\mu$m}, and the total number of elements is in the order of $10^{6}$. The computational platform has an Intel Core i7-7500U processor with 16 GB RAM. The typical run time is $20~$h for the simulation of a streamer propagating for 10 ns, an example of which is presented in Fig. 2.

\begin{figure}[H]
\centering
\subfigure[]{\includegraphics[width=118 bp]{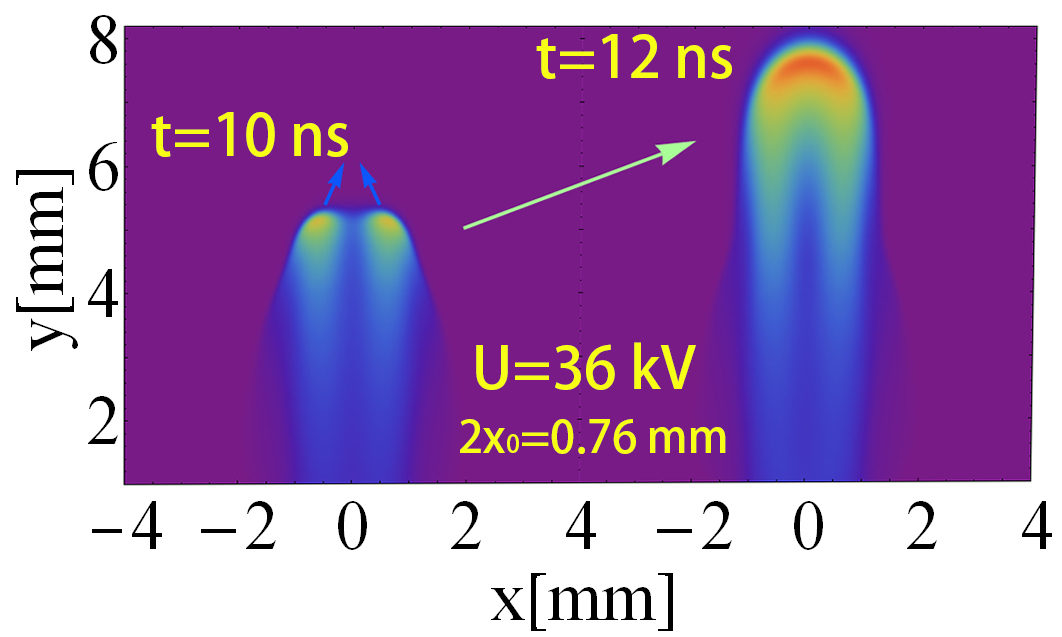}
\includegraphics[height=71 bp]{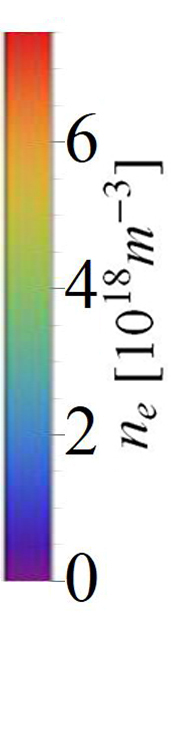}}
\subfigure[]{\includegraphics[width=102 bp]{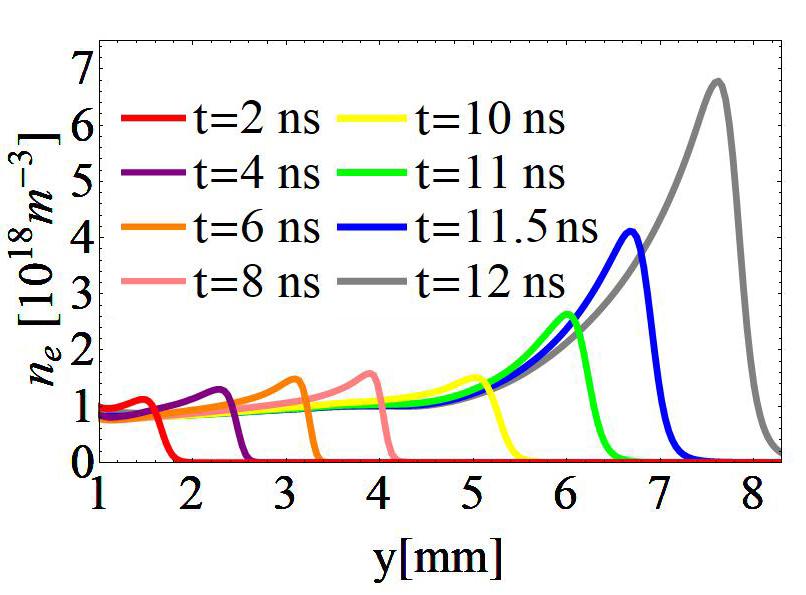}}
\caption{\footnotesize Representative simulation results of streamer discharge with $U=36~$kV and $2x_{0}=0.76~$mm. (a) Electron density distribution at $t=10~$ns when the discharge front has clearly splitted and at $t=12~$ns when the splitted fronts have merged again. (b) Electron density along the $x=0$ line at different instants.}
\end{figure}

In Fig. 2, the average streamer propagation speed (in 12 ns) $v_{ave}=0.64~$m/$\mu$s is within the range of observed results~\cite{Raizer}, and the electron density profiles are consistent with previous studies~\cite{PRE}. From $t=10~$ns to 12 ns, the two splitted discharge fronts are merging (see Fig. 2(a)), which is also reflected in Fig. 2(b) as the rapid increase of electron density at the $x=0$ tip of the streamer after $t=10~$ns. The merging can be attributed to the intensified tip field as the it approaches the cathode (at $y=10$ mm). The increase in the electric field promotes both reactions which contributes to higher electron densities and photoionization which strengthens the tendency of front merging~\cite{Luque2008}. This in turn further increases the electric field at the streamer tip. If the cathode is very far away (keeping background field the same), it would be expected that merging after $t=10~$ns becomes less probable since the tip field does not get such boost from the cathode. In view of this, we limit the scope of this study to the emergence of front splitting while neglecting the subsequent developments, i.e., ending simulation once streamer front advances for $\sim$6 mm or 60\% of the gap length. To probe front splitting, Fig. 2(b) actually gives a hint. Comparing the electron density profiles at $t=8~$ns and $10~$ns, one can see that the peak density decreases by about 10\%, which reverses the trend from $t=2~$ns to 8 ns and is concurrent with the splitting of the front into two heads, each with a center of maximum electron density that moves away from the $x=0$ axis. Therefore, this can be used as a criterion to determine whether and when the front splitting comes into play.

\begin{figure}[H]
\centering
\subfigure{\includegraphics[width=242 bp]{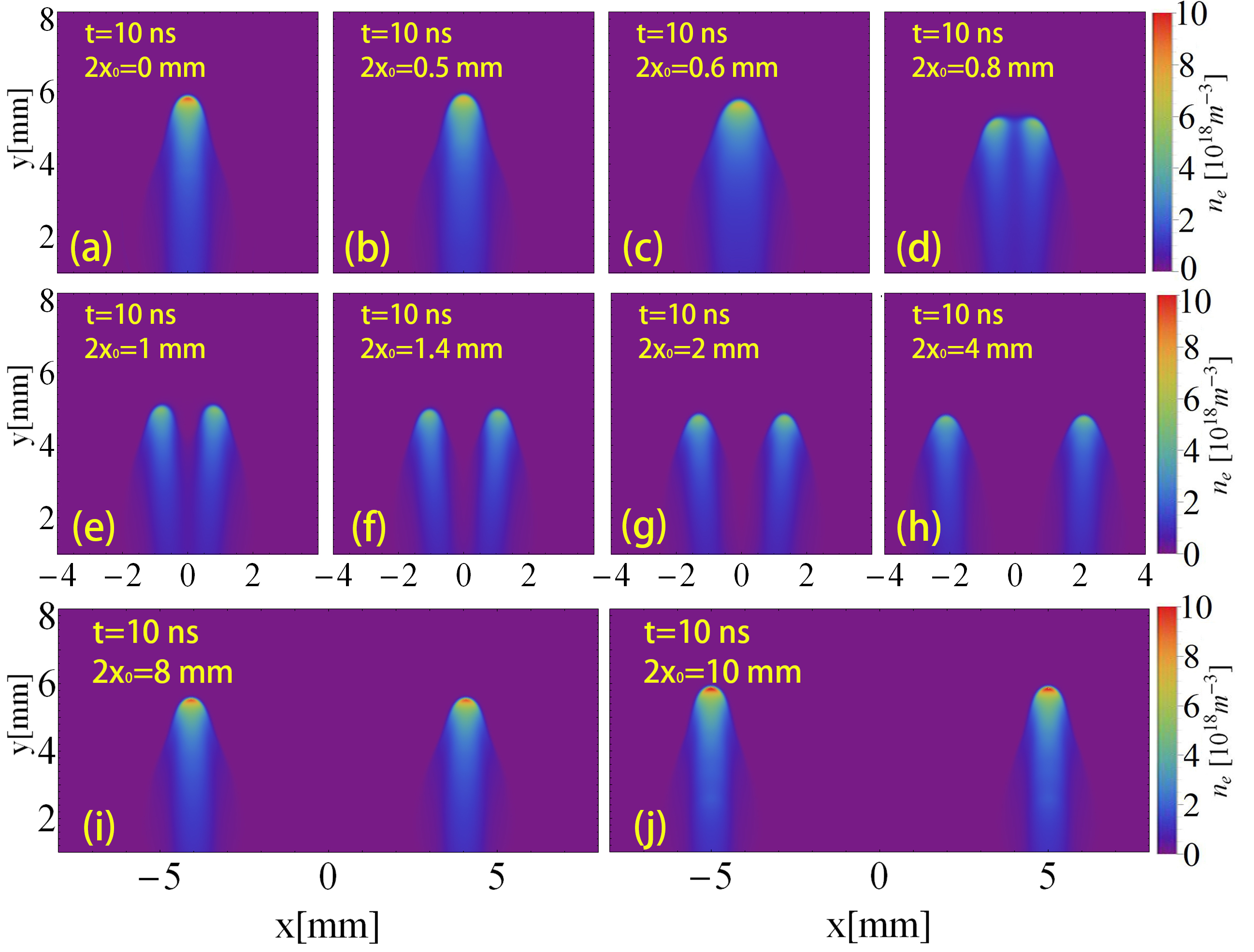}}
\end{figure}
\begin{figure}[H]
\centering
\subfigure{\includegraphics[width=120 bp]{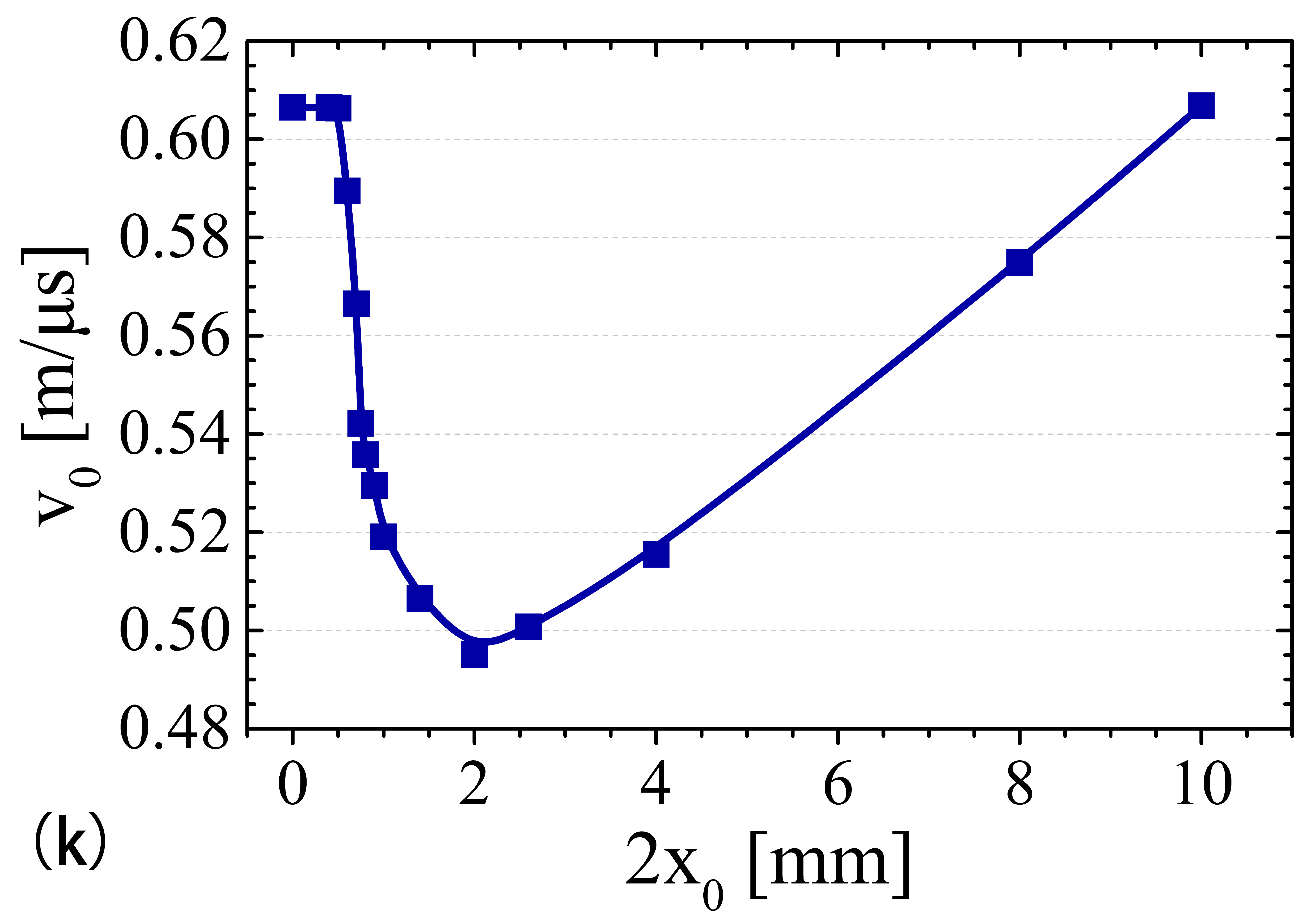}}
\caption{\footnotesize Simulation of streamers with $U=36~$kV and various $2x_{0}$'s from 0 to 10 mm. (a$\sim$j) Electron density distributions at $t=10~$ns. (k) Average streamer speed (in 10 ns) $v_{0}$ vs $2x_{0}$.}
\end{figure}

Fig. 3 aims to show the effect of $2x_{0}$ on streamer dynamics under $U=36~$kV. The electron density distributions at $t=10~$ns in Figs. 3(a)$\sim$(j) indicate that there exists a threshold of $2x_{0}$ below which only one streamer with non-branched front will develop from the superposed Gaussian seeds. The case of $2x_{0}=0$, actually with only one Gaussian seed, serves as the reference group (Fig. 3(a)). When $2x_{0}$ is small (Figs. 3(b,c)), no front splitting occurs, while the streamer spreads more in the $x$ direction. When $2x_{0}$ reaches the threshold (0.76 mm in this case), a lower electron density cleavage emerges near the symmetric axis ($x=0$) of the discharge front as the indicator of front splitting (Fig. 3(d)). When $2x_{0}$ increases to around 1 mm (Fig. 3(e)), the front splitting is more clearly-defined and fully-developed, but extending simulation to 12 ns yields merging results similar to Fig. 2(a). By further increasing $2x_{0}$ to 1.4 mm, the two initial Gaussian seeds are essentially well-separated, which turns the problem into that of two interacting streamers. Due to electrostatic repulsion, the two streamers form an angle as they advance and will not merge at a later time, which resonates with the conclusions in~\cite{Liu2017}. Our result is also in good agreement with the merging criteria proposed in~\cite{Merging2012}, i.e., the ratio between the characteristic width of streamers and their mutual distance needs to be below 0.4 to avoid merging. In Figs. 3(e) and (f), the characteristic width of streamers is 0.8 mm, while the mutual distance is 2.0 mm (former) and 2.4 mm (latter). The ratio is therefore 0.4 in Fig. 3(e) and 0.33 in Fig. 3(f) (even lower in Figs. 3(g)$\sim$(j)). This helps to understand why there is no merging in the latter cases. When $2x_{0}$ exceeds 4 mm (e.g., Fig. 3(h)), the effect of repulsion is no longer obvious, and both streamers propogate in parallel toward the cathode. However, the interaction between streamers tends to slow down the propagation, as can be seen by comparing Fig. 3(h) with Fig. 3(j) in which the interaction is negligible since each of the two streamers has almost the same behavior as the single streamer in Fig. 3(a). This observation is consistent with~\cite{Milikh2016} which did not consider photoionization; so one may conjecture that it is an electrostatic field effect.

The effect of $2x_{0}$ on streamer propagation speed (averaged in the first 10 ns) $v_{0}$ can be found in Fig. 3(k). In addition to the similarity between $2x_{0}=0$ and $2x_{0}=10$ mm cases that we have just seen, there are 3 critical $2x_{0}$ values that demand attention. First, as $2x_{0}$ passes 0.5 mm, the dependence of $v_{0}$ on $2x_{0}$ switches abruptly between insensitive and highly sensitive. Second, at around $2x_{0}$=2 mm, $v_{0}$ hits the lowest point. It is clear that the front splitting threshold (0.76 mm) is neither the $2x_{0}$ corresponding to the extreme speeds nor the $2x_{0}$ at which $v_{0}$ begins to decrease rapidly. And thirdly, the decrease rate of $v_{0}$ starts to decline somewhere between $2x_{0}$=0.7 mm and 0.8 mm, the range in which the front splitting threshold falls. However, with limited data, it is practically difficult to accurately locate this inflection point.

\begin{figure*}
\centering
\subfigure{\includegraphics[width=400 bp]{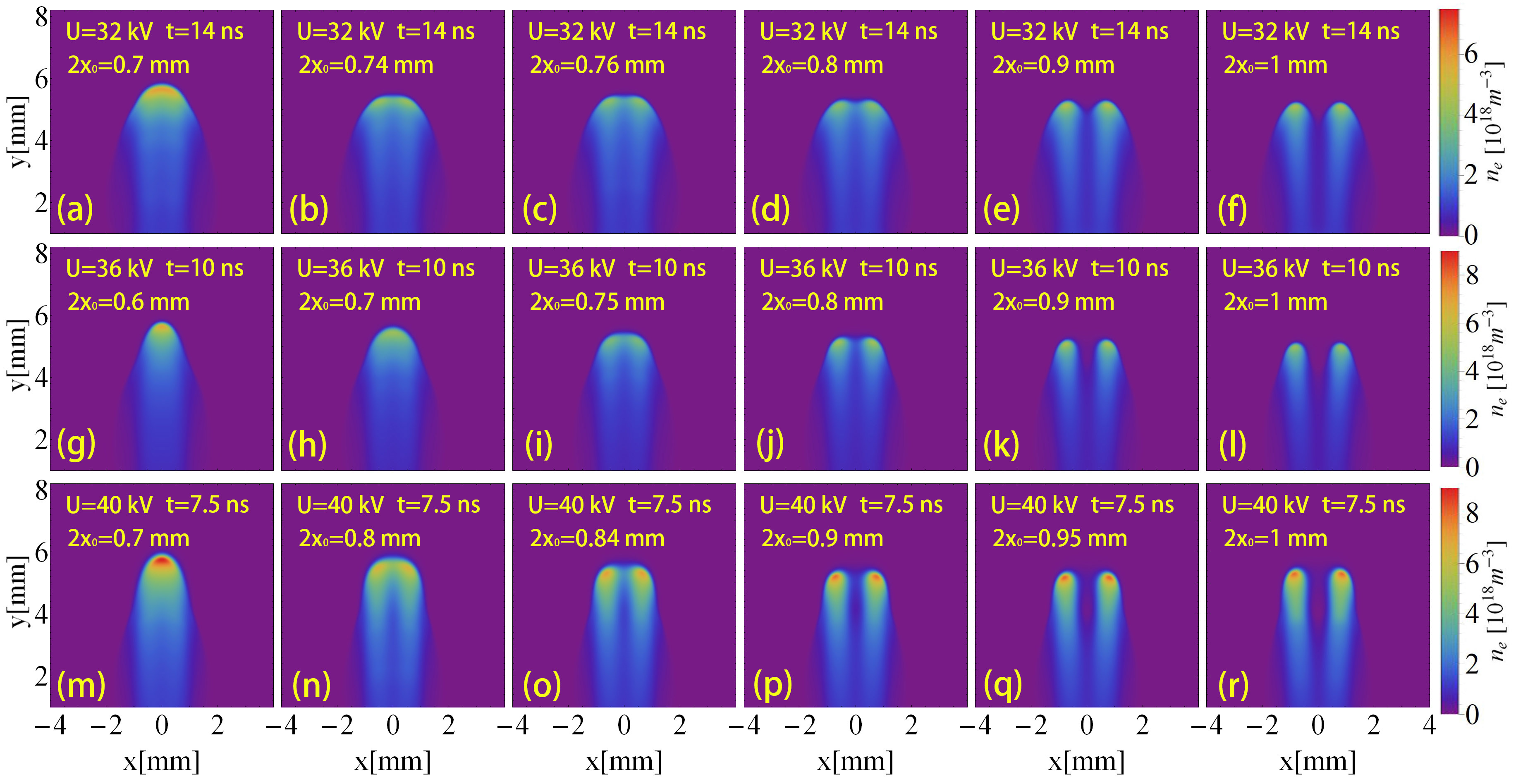}}
\caption{\footnotesize Simulated electron density distributions of streamers with different Gaussian seed distances ($2x_{0}$) under three different voltages: $U=32~$kV (top row, simulation time $14~$ns), $36~$kV (middle row, simulation time $10~$ns), and $40~kV$(bottom row, simulation time $7.5~$ns). The tendency of front spilitting is greater with larger $2x_{0}$'s or under lower voltages. Using the criterion demonstrated earlier in Fig. 2(b), we determined the critical $2x_{0}$'s for front splitting as: $0.74~$mm for $U=32~$kV, $0.76~$mm for $U=36~$kV, and $0.80~$mm for $U=40~$kV.}
\end{figure*}

It is therefore natural to relate the threshold of $2x_{0}$ for front splitting to a better indicator than streamer speed. We have shown in Fig. 1(c) that there are a set of critical values of $2x_{0}$ at which $y_{0}/y_{max}$ drops to 0. These $2x_{0}$'s are solely determined by the ionization seed profiles and adjustable over a fairly wide range by varying electron densities. Using $2x_{0}$ as intermediary, it would be of great interest to establish the connection between seed profile and front splitting. For this purpose, we first extend the streamer simulations in Fig. 3 to more cases with $U=32~$kV and 40 kV, the results of which are presented in Fig. 4. If increasing $2x_{0}$ under the same voltage, similar phenomena to what has been discussed above can be observed. Under higher voltages, the streamer front starts to split at larger $2x_{0}$'s, i.e., the critical values of $2x_{0}$ increases with $U$ ($0.74~$mm for $U=32~$kV, $0.76~$mm for $U=36~$kV, and $0.80~$mm for $U=40~$kV). This indicates that higher voltages or electric fields are more able to hold the streamer front (developed from two overlapped Gaussian seeds) together. A possible explanation is that photoionization is more prominent under higher voltages and suppresses the geometrical irregularity of the front. In addition, higher voltages will speed up the streamer propagation, which is also considered as unfavorable for front splitting or streamer branching.

Now, our attention turns to the quantitative characterization of the ionization seed profiles illustrated in Fig.~1(c). On the horizontal axis of Fig. 1(c), if we locate the 3 critical values of $2x_{0}$ identified in $U=32~$kV, $36~$kV, and $40~$kV cases, the electron densities of the seed profiles corresponding to these critical $2x_{0}$'s are $2.00\times 10^{19}~$ \text{$m^{-3}$}, $1.88\times 10^{19}~$ \text{$ m^{-3}$}, and $1.76\times 10^{19}~$ \text{$m^{-3}$}, respectively. Hereafter, the seed profiles associated with these specific electron densities will be referred to as indicative profiles. In other words, one may observe the change in the shape of the indicative profiles, as the Gaussian seed distance $2x_{0}$ increases from 0. Once $2x_{0}$ arrives at the corresponding critical value, the indicative profile turns into two semicircles and the front starts splitting into two lobes. In this way we have related the geometric irregularity of ionization seed profile to the front splitting of streamer discharge. The higher the applied voltage is, the lower the electron density of the indicative profile would be. This accords with the physics intuition that applied voltage (or electric field) and preionization level, within certain limits, are two complementary factors in determining the dynamics of streamer discharge.

A further discussion on the applicability or valid parameter range of the indicative profile approach is given as follows. First, we consider the cases with various characteristic sizes ($\sigma_{x}=\sigma_{y}=\sigma$) of the ionization seed. Since no change is made to the preionization level $n_{0}$, it is expected that the same set of electron densities of indicative profiles as those identified in the previous paragraph will also apply here. Note that after changing $\sigma$, $y_{0}/y_{max}$ as functions of $2x_{0}$ will be different, even for the same electron density. The results in Fig. 5 (left column) clearly show that with the increase of $\sigma$, the indicative profiles as well as the critical values of $2x_{0}$ for front splitting shift to the right, i.e., the two Gaussian seeds need to be separated at larger distances for front splitting to occur. Then extensive simulations are run to check the validity/accuracy of the above method. In these cases we keep $2x_{0}=1~$mm and choose $\sigma$ in the range of 0.45$\sim$0.65 mm (increment 0.02 mm). Selected results of the simulated electron density distributions are presented in Fig. 5. It is discovered that when $\sigma$ is below 0.55 mm, under all three voltages there is front splitting; when $\sigma$ is $0.55~$mm and $0.57~$mm, front splitting happens under two lower voltages; when $\sigma$ is $0.59~$mm and $0.61~$mm, front splits only under $32~$kV; when $\sigma$ is above $0.61~$mm, there is no front splitting at all.

\begin{figure}[H]
\centering
\subfigure{\includegraphics[width=260 bp]{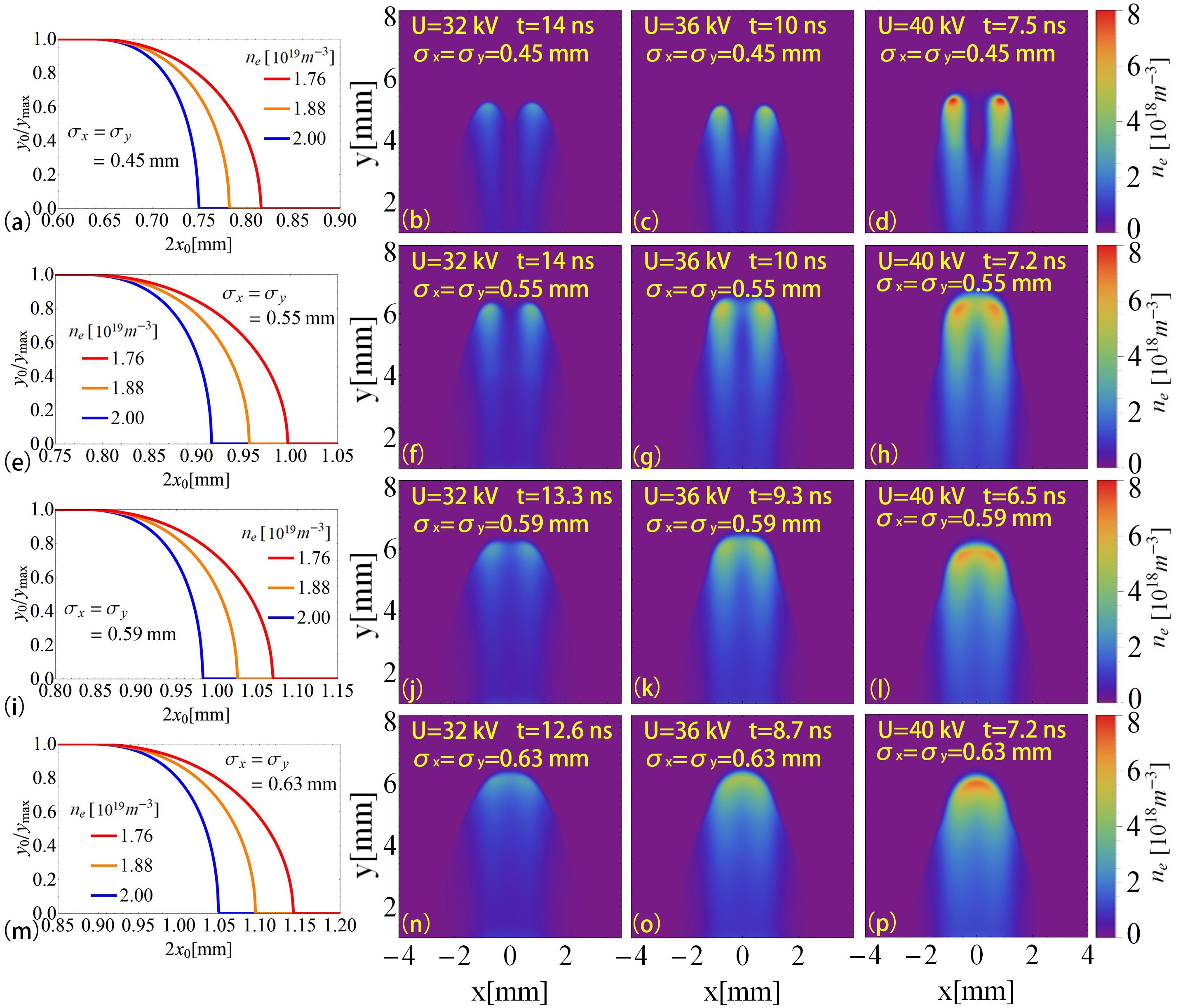}}
\caption{\footnotesize Simulation of streamers with different characteristic sizes of Gaussian seed, $\sigma=0.45~$mm (Row 1 from top), $0.55~$mm (Row 2), $0.59~$mm (Row 3), and $0.63~$mm (Row 4). The 1st column [(a)(e)(i)(m)] presents quantitative characterization of indicative profiles for $U=32~$kV, $36~$kV, and $40~$kV. The critical $2x_{0}$'s for front splitting are at the intersections of these curves with the horizontal axis. The 2nd to 4th columns are the electron density distributions of streamers with $2x_{0}=1~$mm under 3 voltages.}
\end{figure}

As show in Fig. 5, when $\sigma=0.45~$mm, under all three voltages, $2x_{0}=1~$mm exceeds the critical value for front splitting (Fig.~5(a)), which is cross-checked by the splitted streamers in Figs. 5(b)$\sim$(d). At $\sigma=0.55~$mm, $2x_{0}=1~$mm is greater than the critical values under 32 kV and 36 kV and very close to the one under 40 kV (Fig.~5(e)). In Fig. 5(h), the discharge front of the streamer is not as clearly splitted under $40~$kV as those under $32~$kV and $36~$kV (Figs. 5(f) and (g)). In Fig.~5(i), when $\sigma=0.59~$mm,  $2x_{0}=1~$mm is only above the critical value under $32~$kV. Correspondingly, only the streamer under 32 kV has front splitting (Fig. 5(j)), while the other two maintain an integral front (Figs. 5(k) and (l)). In the case of $\sigma=0.63~$mm, $2x_{0}=1~$mm falls below all three critical values (Fig. 5(m)) and there is no front splitting for any of the three streamers (Figs. 5(n)$\sim$(p)). Moreover, we find out that under the same voltage, there is a simple linear relationship between the critical $2x_{0}$ and $\sigma$ (at least in the range of $0.45$$\sim$$0.65~$mm). For example, under $40~$kV ($32~$kV), the ratio of critical $2x_{0}$ to $\sigma$ is 0.8/0.45$\approx$1.8 (0.74/0.45$\approx$1.6). Now that the actual $2x_{0}$=1 mm, then $\sigma$ should not exceed 1/1.8$\approx$0.56 mm ($1/1.6$$\approx$$0.62~$mm) for the emergence of front splitting. These estimations are consistent with the results in Fig. 5. The minor discrepancy may have resulted from the relatively large increment ($0.02~$mm) when scanning the range of $\sigma$. Therefore, the determination of critical values of $2x_{0}$ from indicative profiles can serve as a consistent and simple tool for the prediction of front splitting in streamer discharges. In principle, this conclusion holds for various preionization levels or $n_{0}$'s (as long as the streamer simulation still has good convergence and stability). It is needed to select different sets of indicative profiles associated with appropriate electron densities.

So far we have only discussed the cases with two identical Gaussian seeds. There would be two challenges when applying the approach demonstrated before to the cases of two Gaussian seeds with either different $\sigma$'s or different $n_{0}$'s. Firstly, if one seed has a $\sigma$ or $n_{0}$ that is significantly higher ($>$ 1.5 times) than the other, our simulations indicate that the streamer developed from this seed will dominate the process; in most cases the weaker streamer would simply become part of the dominate one during very early stage of development ($<$ 2 ns). No front splitting occurs. Secondly, even if the values of $\sigma$ and $n_{0}$ of the two seeds are close, due to the asymmetry, we cannot define $y_{0}/y_{max}$ as easily as in Fig. 1(b). Hypothetically, one can still try some alternative measures of the ionization seed profiles in certain scenarios, which, however, is beyond the scope of this study.

In summary, we have performed 2D simulations of negative streamers in air using COMSOL Multiphysics with a focus on the effect of the geometrical irregularity of the ionization seed on the front splitting of the streamer discharge. Specifically, we consider the cases in which the ionization seed can be modeled as the superposition of two identical Gaussians separated at distance $2x_{0}$. When the two Gaussians are heavily overlapped, we use a ratio $y_{0}/y_{max}$ to quantitatively characterize the geometrical irregularity of the ionization seed profiles (In an approximate sense, 1 means one semicircle, 0 means two separate semicicles, while the ratio between 1 and 0 means seed profiles with large geometrical deviations from regular semicircles). By equating the critical $2x_{0}$ for front splitting to the $2x_{0}$ at which $y_{0}/y_{max}$ just drops to 0, one can identify indicative profiles with corresponding electron densities. These indicative profiles can be applied to cases with various characteristic seed sizes ($\sigma$) and preionizaion levels ($n_{0}$) to predict if front splitting is likely to occur. Although it is well-accepted that irregularly-shaped discharge fronts and ionization seeds may cause branching, this work makes significant contributions to the field via a systematic and quantitative investigation of geometrical irregularity, front splitting, and their connections. More numerical studies can be conducted in 3D to further verify our conclusions. It is also interesting to see the differences made after stochastic factors are added to the model, which might inform the modification of the indicative profiles method to take into account randomness in the process. On the other hand, interestingly, our work may also have experimental implications. Assuming the technology is ready for the detection and imaging of ionization seed profiles corresponding to certain electron densities, and specifically-designed electrode surfaces are used to create desired ionization seeds, we would be able to not only identify and test the streamer branching mechanisms, but also reliably predict and even control the streamer front splitting for some important applications such as material processing.

\section*{Acknowledgment}
This research has been funded by the National Natural Science Foundation of China under grant 51577098.


\begin{thebibliography}{99}
\bibitem{Raizer}
Y.\ P.\ Raizer, \textit{Gas Discharge Physics} (Springer-Verlag, Berlin, 1991).

\bibitem{Winands}
G.\ Winands, K.\ Yan, A.\ Pemen, S.\ Nair, Z.\ Liu, and E.\ van Heesch, IEEE Trans.\ Plasma Sci. \textbf{34}, 2426 (2006).

\bibitem{Oehrlein}
G.\ S.\ Oehrlein and S.\ Hamaguchi, Plasma Sources Sci.\ Technol. \textbf{27}, 023001 (2018).

\bibitem{Fridman}
G.\ Fridman, G.\ Friedman, A.\ Gutsol, A.\ B.\ Shekhter, V.\ N.\ Vasilets, and A. Fridman, Plasma Process.\ Polym. \textbf{5}, 503 (2008).

\bibitem{Moreau}
E.\ Moreau, J.\ Phys.\ D: Appl.\ Phys. \textbf{40}, 605 (2007).

\bibitem{Starikovskiy}
A.\ Starikovskiy and N.\ Aleksandrov, Prog.\ Energy Combust.\ Sci. \textbf{39}, 61 (2013).

\bibitem{Dubinova}
A.\ Dobinova, C.\ Rutjes, U.\ Ebert, S.\ Buitink, O.\ Scholten, and G.\ T.\ N.\ Trinh, Phys.\ Rev.\ Lett. \textbf{115}, 015002 (2015).

\bibitem{Zhang}
X.\ Zhang, Y.\ Zhu, S.\ Gu, and J.\ He, Appl.\ Phys.\ Lett. \textbf{111}, 224101 (2017).

\bibitem{Shneider1}
Y.\ P.\ Raizer, G.\ M.\ Milikh, and M.\ N.\ Shneider, J.\ Geophys.\ Res. \textbf{115}, A00E42 (2010).

\bibitem{Shneider2}
N.\ A.\ Popov, G.\ M.\ Milikh, and M.\ N.\ Shneider, J.\ Atmos.\ Sol.-Terr.\ Phys. \textbf{147}, 121 (2016).

\bibitem{Nijdam2010}
S.\ Nijdam, F.\ M.\ J.\ H.\ van de Wetering, R.\ Blanc, E.\ M.\ van Veldhuizen, and U.\ Ebert, J.\ Phys.\ D: Appl.\ Phys. \textbf{43}, 145204 (2010).

\bibitem{Nijdam2016}
S.\ Nijdam, J.\ Teunissen, E.\ Takahashi, and U.\ Ebert, Plasma Sources Sci.\ Technol. \textbf{25}, 044001 (2016).

\bibitem{Liu2012}
N.\ Liu, B.\ Kosar, S.\ Sadighi, J.\ R.\ Dwyer, and H.\ K.\ Rassoul, Phys.\ Rev.\ Lett. \textbf{109}, 025002 (2012).

\bibitem{Heijmans2013}
L.\ C.\ J.\ Heijmans, S.\ Nijdam, E.\ M.\ van Veldhuizen, and U.\ Ebert, Europhys.\ Lett. \textbf{103}, 25002 (2013).

\bibitem{Nijdam2014}
S.\ Nijdam, E.\ Takahashi, J.\ Teunissen, and U.\ Ebert, New\ J.\ Phys. \textbf{16}, 103038 (2014).

\bibitem{Zhuang1}
C.\ Zhuang, and R.\ Zeng, Commun.\ Comput.\ Phys. \textbf{15}, 153 (2014).

\bibitem{Zhuang2}
C.\ Zhuang, and R.\ Zeng, Appl.\ Math.\ Comput. \textbf{219}, 9925 (2013).

\bibitem{Dujko2013}
S.\ Dujko, A.\ H.\ Markosyan, R.\ D.\ White, and U.\ Ebert, J.\ Phys.\ D: Appl.\ Phys. \textbf{46}, 475202 (2013).

\bibitem{Luque2014}
A.\ Luque, and U.\ Ebert, New\ J.\ Phys. \textbf{16}, 013039 (2014).

\bibitem{Teunissen2017}
J.\ Teunissen and U.\ Ebert, J.\ Phys.\ D:\ Appl.\ Phys. \textbf{50}, 474001 (2017).

\bibitem{Luque2008}
A.\ Luque, U.\ Ebert, and W.\ Hundsdorfer, Phys.\ Rev.\ Lett. \textbf{101}, 075005 (2008).

\bibitem{Merging2012}
Z.\ Bonaventura, M.\ Duarte, A.\ Bourdon, and M.\ Massot, Plasma Sources Sci.\ Technol. \textbf{21}, 052001 (2012).

\bibitem{Milikh2016}
G.\ M.\ Milikh, A.\ V.\ Likhanskii, M.\ N.\ Shneider, and A.\ George, J.\ Plasma\ Phys. \textbf{82}, 905820102 (2016).

\bibitem{Liu2017}
F.\ Shi, N.\ Liu, and J.\ R.\ Dwyer, J.\ Geophys.\ Res.:\ Atmos. \textbf{122}, 10169 (2017).

\bibitem{Levko2017}
D.\ Levko, M.\ Pachuilo, and L.\ L.\ Raja, J.\ Phys.\ D:\ Appl.\ Phys. \textbf{50}, 354004 (2017).

\bibitem{DBM}
L.\ Niemeyer, L.\ Pietronero, and H.\ J.\ Wiesmann, Phys.\ Rev.\ Lett. \textbf{52}, 1033 (1984).

\bibitem{Zhang1}
X.\ Zhang, L.\ Dong, J.\ He, S.\ Chen, and R.\ Zeng, J.\ Lightning Res. \textbf{1}, 1 (2009).

\bibitem{Zhang2}
J.\ He, X.\ Zhang, L.\ Dong, R.\ Zeng, and Z.\ Liu, Sci.\ China E: Technol.\ Sci. \textbf{52}, 3135 (2009).

\bibitem{MA2002}
M.\ Arrayas, U.\ Ebert, and W.\ Hundsdorfer, Phys.\ Rev.\ Lett. \textbf{88}, 174502 (2002).

\bibitem{MA2005}
M.\ Arrayas, M.\ A.\ Fontelos, and J.\ L.\ Trueba, Phys.\ Rev.\ Lett. \textbf{95}, 165001 (2005).

\bibitem{Ebert2008}
G.\ Derks, U.\ Ebert, and B.\ Meulenbroek, J.\ Nonlinear Sci. \textbf{18}, 551 (2008).

\bibitem{RU2013}
L.\ A.\ Savel'eva, A.\ V.\ Samusenko, and Y.\ K.\ Stishkov, Surf.\ Eng.\ Appl.\ Electrochem. \textbf{49}, 125 (2013).

\bibitem{Liu2015}
S.\ Sadighi, N.\ Liu, J.\ R.\ Dwyer, and H.\ K.\ Rassoul, J.\ Geophys.\ Res.\ Atmos. \textbf{120}, 3660 (2015).

\bibitem{Papageorgiou}
L.\ Papageorgiou, A.\ C.\ Metaxas, and G.\ E.\ Georghiou, J.\ Phys.\ D: Appl.\ Phys. \textbf{44}, 045203 (2011).

\bibitem{pre84}
A.\ Luque, and U.\ Ebert, Phys.\ Rev.\ E \textbf{84}, 046411 (2011).

\bibitem{Li}
C.\ Li, J.\ Teunissen, M.\ Nool, W.\ Hundsdorfer, and U.\ Ebert, Plasma Sources Sci.\ Technol. \textbf{21}, 055019 (2012).

\bibitem{TE2016}
J.\ Teunissen, and U.\ Ebert, Plasma Sources Sci.\ Technol. \textbf{25}, 044005 (2016).

\bibitem{probing}
G.\ Wormeester, S.\ Pancheshnyi, A.\ Luque, S.\ Nijdam, and U.\ Ebert, J.\ Phys.\ D: Appl.\ Phys. \textbf{43}, 505201 (2010).

\bibitem{comparing}
A.\ H.\ Markosyan, J.\ Teunissen, S.\ Dujko, and U.\ Ebert, Plasma\ Sources Sci.\ Technol. \textbf{24}, 065002 (2015).


%
%
%

%
%
%
%
%
%
%
%
%
%
%
%
%
\bibitem{PRE}
S.\ Pancheshnyi, M.\ Nudnova, and A.\ Starikovskii, Phys.\ Rev.\ E \textbf{71}, 016407 (2005).

\bibitem{17}
A.\ Bourdon, N.\ P.\ Pasko, N.\ Y.\ Liu, S.\ C\'{e}lestin, P.\ S\'{e}gur, and E.\ Marode,  Plasma Sources Sci.\ Technol.\ \textbf{16}, 656 (2007).

\bibitem{PIC}
D.\ Breden, K.\ Miki, and L.\ L.\ Raja, Plasma Source Sci.\ Technol. \textbf{21}, 034011 (2012).

%


%
%
%
%
%
%
%
%
%
%
%
%
%
%
%
%
%
%
%
%
%
%
%
%
%
%
%
%
%
%
%
%
%
%
%
%
%
%
%
%
%
%
%
%
%
%
%
%
%
%
%
%
%
%
%
%
%
%
%
%
%
%
%
%
%
%
%
%
%
%
%
%
%
%
%
%
%
%
%
%
%
%
%

\end{thebibliography}
\end{document}